# A Practical Localization Algorithm Based on Wireless Sensor Networks

Tao Huang, Zhikui Chen, Feng Xia, Cheng Jin, Liang Li
School of Software
Dalian University of Technology
Dalian 116620, China
ht2411@hotmail.com, zkchen@dlut.edu.cn, f.xia@ieee.org, {jincheng117, liangli0917}@gmail.com

*Abstract*—Many localization algorithms and systems have been developed by means of wireless sensor networks for both indoor and outdoor environments. To achieve higher localization accuracy, extra hardware equipments are utilized by most of the existing localization algorithms, which increase the cost and greatly limit the range of location-based applications. In this paper we present a method which can effectively meet different localization accuracy requirements of most indoor and outdoor location services in realistic applications. Our algorithm is composed of two phases: partition phase, in which the target region is split into small grids and localization refinement phase in which a higher accuracy location can be generated by applying a trick algorithm. A realistic demo system using our algorithm has been developed to illustrate its feasibility and availability. The results show that our algorithm can improve the localization accuracy.

## I. INTRODUCTION

Wireless sensor networks (WSNs) [1] has been regarded as a promising data acquisition tool, which has been gradually used in industrial filed as well as national defense field. As a cross-discipline, it integrates micro-electrical technique, short-range communication technique and embedded system technique into together. However, big challenge has been faced as its stringent constraints in energy consumption, memory storage and communication bandwidth. In spite of this, as a supporting technique in WSNs, localization problem has been given a great concern, especially in data marker, geographic routing and data aggregation algorithm, where data are meaningless if no location information collected.

Traditional location technique such as GPS cannot be directly used in WSNs, as its costly requirement of sophisticated equipments and higher energy consumption in order to ranging, which have greatly constrained the application scale of WSNs. In WSNs energy conservation has been considered as the core issue and the costs as well as size of a node should be as small as possible, to apply into large-scale applications for a long time. To address this issue, many localization algorithms developed did not use GPS technique directly, but employ it as assistance one in some case, and more efforts are focus on the excavation of WSNs itself. Till now, all the algorithms proposed can be broadly classified into two categories: range-based localization [3, 4, 5, 6, 7] and range-free localization [9, 10, 11, 12, 13, 14]. The former technique usually needs extra hardware to accomplish ranging, and then utilize some algorithm to calculate position coordinate. The latter technique exploits the characteristics of network connectivity [9, 10, 11] (such as DV-hop [9], MDS-MAP [10], Amorphous [11] etc.), proximity information [12, 13, 14] (such as APIT [12], Centroid [13], Convex [14], etc.) to realize a tolerable position estimation. Although it cannot provide accuracy compared with range-based scheme, it holds a cost-effective advantage, which is an enormous attraction for large-scale system deployment.

The remainders of this paper are organized as follows: section 2 surveys related work. Our algorithm is explained in section 3. Section 4 presents a demo system and our experiment work. Finally, section 5 concludes this paper.

## II. RELATED WORK

Many localization algorithms have been developed in WSNs, all of which can be roughly categorized into one of the follows: range-based localization [3, 4, 5, 6, 7] and range-free localization [9, 10, 11, 12, 13, 14]. Range-based localization always has two phases to go: ranging and position computation. In the first phase it utilize some ranging method such as TOA (Time of Arrival) [4], TDOA (Time Difference of Arrival) [5], AOA (Angle of Arrival) [6] and RSSI (received signal strength indicator) [7] to obtain the distance between two nodes (always blind node whose position unknown and reference nodes also called beacon nodes whose position pre-known). With the coordinate knowledge of the reference node attached with RSSI, blind node (also refers to the mobile node or target) can calculate its own coordinate by using some methods, such as Trilateration, Triangulation, and Maximum likelihood estimation.

In TOA based localization system, it requires extra hardware to guarantee the synchronization between transmitting equipment and receiving equipment, otherwise, a small timing error may result in tremendous distance estimation error. In TDOA systems, it shares the same drawbacks with the TOA systems where they call for expensive hardware. Moreover, TDOA employs ultrasound ranging technique, which needs density deployment as the transmission distance of ultrasound is merely 20-30 feet. AOA localization system can be regarded as complementary technique for TOA and TDOA. It allows nodes to estimate the distance according to relative angles, which can be achieved by installing angle measuring equipment. For this point, it is also not advised to be used in large scale sensor networks. On the contrary, RSSI technique overcomes a majority of shortcomings mentioned above. It utilizes some signal propagation models, either from theoretical or empirical, to translate signal strength into distance, thus it doesn't need additional hardware. As everything has two sides, this technique usually suffers from multi-path fading, noise interference, and irregular signal propagation, which has severely affected the accuracy of ranging estimate. Although existence of these disadvantage, we still can alleviate this suffering in some special method, and we argue that only if proper measures are been taken, the localization accuracy can be promoted to meet most of application systems. We achieve this by employing regular deployment of node, region partition and localization refinement.

In RSSI localization systems, distance estimation between transmitter and receiver by using received signal strength on some signal propagation model should be accomplished previously. The widely used propagation model is log-normal shadowing model expressed as:

$$P_r(d)[dBm] = P_r(d_0)[dBm] - 10n\log_{10}(d/d_0) + x_\sigma[dBm] \quad (1)$$

Where $d$ is the distance between transmitter and receiver, $d_0$ reference distance, $P_r(d)$ the received power, $P_r(d_0)$ the received power of the point with a reference distance $d_0$, n path loss exponent factor which is related to environment, $x_\sigma$ Gaussian random variable which reflect the change of power when distance is fixed.

However, in practical we use the simplified shadowing model:

$$P_r(d)[dBm] = P_r(d_0)[dBm] - 10n\lg(d/d_0) \quad (2)$$

Usually, we select $d_0$ as 1 meter, so we have:

$$RSS[dBm] = P_r(d)[dBm] = A - 10n\lg d \quad (3)$$

Where $A$ is the received signal power of receiver from a transmitter one meter away.

We carry out system by using CC2430 chip, which is system-on-chip solution for 2.4GHz IEEE 802.15.4/Zigbee with the characteristics of low-power and low date rate (up to 250kbps). CC2430 has a build-in register called RSSI_VAL for storing RSSI value, and the power value on RF pin is:

$$RSS = RSSI\_VAL + RSSI\_OFFSET[dBm] \quad (4)$$

Empirically, the $RSSI\_OFFSET$ can be assigned -45dBm. Up to now, we can estimate the distance between the transmitter and receiver.

### III. LOCALIZATION PROCESSING

In this section we will present our localization algorithm, which consists of two phases: region partition and localization refinement. The brief principal of this method is that we first split the target region into small grids by deploying sensor nodes regularly, and the nodes reside on the vertex of the grid. The grid that the blind nodes current located in can be easily determined by comparing their RSSI strength as the distance is shorter the RSSI strength is stronger and vice versa. Next in this determined grid we refine the position coordinate to meet our accuracy requirement by employ a trick algorithm developed by us.

### A. Region Partition

To simplify this problem we consider dividing the target region into grid (square or rectangle) (Fig. 1). Firstly setup the coordinate XOY for our target region. And then draw lines which parallel to the x and y axis respectively. The feature of the line must be paralleling to coordinate axis have special meanings which we will explain in section 3.2. Actually the distance between any two lines are not constrained as a constant number, that's to say, you can designed the distance you want between two lines neighbored according to the accuracy requirement of your system. Usually the distance can be 5 to 10 meters approximately, which is not a restriction of using our algorithm. If you want improve the system precision you can set, for example 5 meters or less, and if the precision requirement of your system is not so high, you can set it 50 meters or even 100 meters as long as the distance of this two nodes are not beyond the radio communication radius.

Next, we determine which grid the blind node belongs to, but before that we assume that the RSSI value can correctly reflect the distance that a blind node is near or far to a reference node. When the blind node moves into a grid $S$, it broadcasts message, any reference nodes which receive this message can extract a RSSI value from it. After a short time (usually less than 200 milliseconds) accumulation of RSSI values followed by average value calculation, a RSSI value is generated and returned to blind node. For a blind node, it may receive many RSSI values from its surroundings after broadcast, however we just pick out the maximum four values with its corresponding coordinator as $(x_1, y_1)$, $(x_2, y_2)$, $(x_3, y_3)$, $(x_4, y_4)$, the reason comes from our experimental result that with the distance increasing the error of distance estimation also ascend and hard to describe its rule, which is depicted in Figure 4. From above theoretically we can easily infer that this coordinate set can form a rectangle, in which the blind node resides. However, in practice the coordinate set may not be able to construct a rectangle when blind node move from one grid to another or for the sake of broadcast period is too long. So firstly we should validate that whether the coordinate set can form a rectangle, if so, we can safely conclude that blind node are located in this rectangle. Otherwise, Fig. 2 depicts this circumstance. (1) is the case blind node move from one grid to another, and (2) is the case blind node approach to a reference node. The common characteristic of these two cases is that the four reference nodes $A, B, C, D$ from which blind node receive four maximum RSSI values cannot form a rectangle. For the former case, we divide these four points into two groups, in each of which the difference value of two RSSI is minimal, for instance $A, B$ group and $C, D$ group. Then calculate the x-axis value and y-axis value respectively, and the two parts make up the position coordinate of blind node. For the latter case, by using the maximum RSSI we can calculate the distance r (this will be introduced in section 2) to one reference node, such as $A$ in Fig. 2 (2). We can say that the blind node is in the circle with $A$ as its center and radius is r. But how to calculate its direction, that's to say, how can we know which grid the bind node reside in now. To solve this issue, we just utilize some variables to memory the former moment which grid the blind node belongs to, and employ them to infer the current direction information.

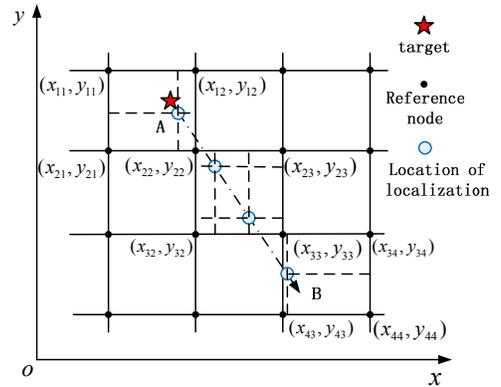

Figure 1. Partition the region into grids, in which the blind node can locate.

Till now, we can determine which grid blind node are locate in currently, and then we can constrain the localization error within the bound of half length of the grid edge, by comparing the four maximum RSSI values received, and their corresponding reference nodes must be on the grid vertices. For example, in Fig. 3, if the $RSSI_A > RSSI_B$ and $RSSI_D > RSSI_C$ we can infer that blind node is at the left side of this grid; if $RSSI_D > RSSI_A$ and $RSSI_C > RSSI_B$ we know that blind node is at the top side of this grid.

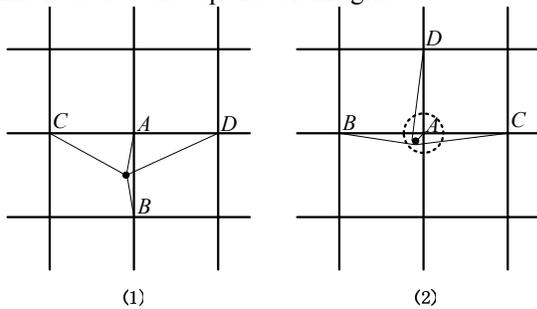

Figure 2. Coordinate set cannot form a rectangle.

### B. Localization Refinement

From 3.1 a rough position of blind node can be obtained, and this may satisfy the requirement for part of applications where it needs only approximate location or the situation when the density of the reference nodes deployed is high enough. However, by increasing the node density to gain a high precision is not a wise selection as it costly requirement for considerable reference nodes, which has become an obstacle for large-scale applications. To overcome this problem, we accomplish a fine-grain localization method for blind node in a grid, as depicted in Fig. 3, by using a compact algorithm which can easily implement two-dimensional plane localization with a regular deployment of reference nodes.

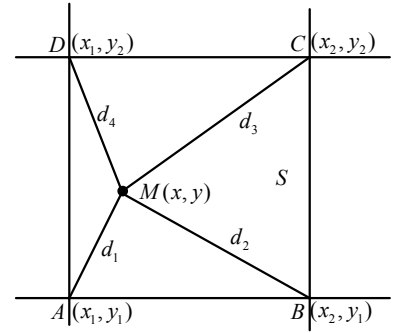

Figure 3. Position calculation of blind node in a grid.

The requirement of this algorithm is that the reference node must be deployed along with the axis (x-axis and y-axis). In Fig. 3, explore $\triangle AMB$ and $\triangle CMD$ we have equations set:

$$\begin{cases} d_1^2 - (x - x_1)^2 = d_2^2 - (x_2 - x)^2 \\ d_4^2 - (x - x_1)^2 = d_3^2 - (x_2 - x)^2 \end{cases} \quad (5)$$

In practice, usually this equations set has no solutions as the ranging error is different for the four nodes. However, by each equation of this equations set we can calculate an x-coordinate value, so our solution is as follows: calculate two x-coordinate values respectively by using equations set (5), and then calculate the average of these two values as the final x-coordinate value for blind node, which is:

$$x = \frac{1}{2} \cdot [x_1 + x_2 - \frac{(d_1^2 + d_4^2) - (d_2^2 + d_3^2)}{2(x_1 - x_2)}] \quad (6)$$

Using the same philosophy, we can easily obtain the y-coordinate value:

$$y = \frac{1}{2} \cdot [y_1 + y_2 - \frac{(d_1^2 + d_2^2) - (d_3^2 + d_4^2)}{2(y_1 - y_2)}] \quad (7)$$

Therefore, the final coordinate of blind node consists of (6) (7).

## IV. EVALUATION

We have implemented our algorithm in a realistic environment. The chip we used is cc2430 provided by TI Corporation, and it is a system-on-chip solution for short distance wireless communication applications. It has integrated RF components, which provide RSSI information with ability of low energy consumption, thus it can help us construct our localization application easily. Also, we run a Zigbee stack named z-stack (version: 2.4.3) made by TI Corporation on this chip platform. Although a localization system has existed and completed by them using the CC2431 chip, the function of which almost equivalent to the CC2430 except for owing a localization engine. The localization error is within 3 meters which has been declared in their explanation documents. In addition, we found that when the blind node moves near to the reference nodes, localization errors usually larger than in the other region, like the center of the region according to our realistic environment test.

First, we show the ranging results and their errors we have tested at the indoor and outdoor environment, as illustrated in Fig. 4. In Fig. 4 (1) we can see that the calculated distance in outdoor environment are more concentrate and close to the real distance, while in indoor environment are more scattered and exist many irregular nodes with the distance ascend. The same point is that the shorter real distance is, more concentrate calculated distance we will get. (2) Shows that the average error of different real distance in outdoor and indoor environment, from which we can see that in our environment real distance less than four or five meters can obtain a good ranging result. Thus in our localization system we set the distance between any two reference nodes four meters.

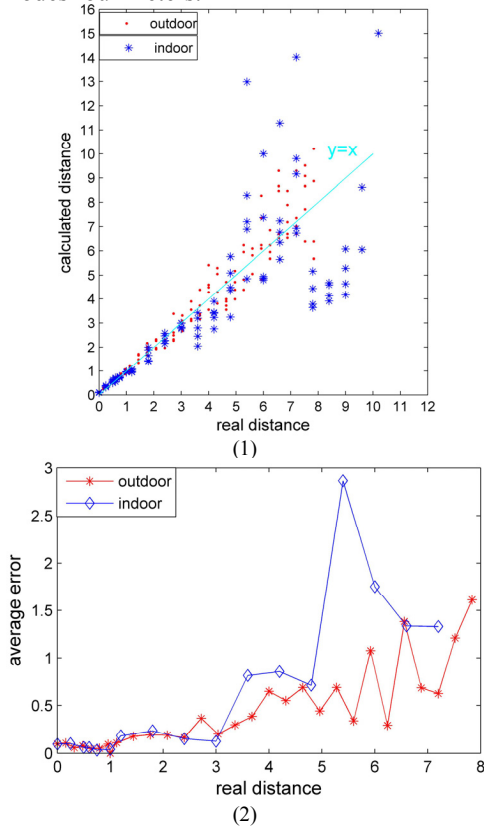

Figure 4. Distance estimation and errors in different environment: (1) Real distance and calculated distance in outdoor and indoor environment. (2) Average calculated error at different distance in outdoor and indoor environment.

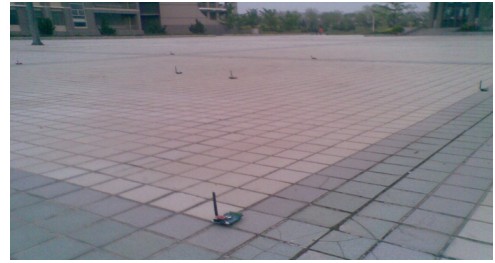

(1)

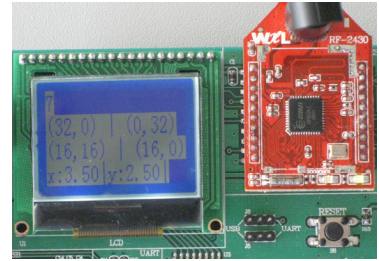

(2)

Figure 5. Our test environment and sensor node.

To illustrate the performance of our system, we compare it with the CC2431 localization system developed by TI Corporation. Fig. 5 (1) shows our outdoor test environment and (2) is the real node we use. Both these two systems deployed at the same place nearly the same period time, with the same spacing distance (four meters) of reference nodes. Also we try our best to make the other environment related differences of these two systems at running time are as small as possible. We sampled 625 data in 8m*8m square area. Fig. 6 shows the 3D localization results and errors comparison in different points. The z-axis shows the error value, which represented the distance between calculated coordinate (x', y') and real coordinate (x, y). On the whole, our system in Fig. 6 (2) harvest a more accurate localization results compared with CC2431 system in Fig. 6 (1), on both the edge of two reference nodes and the area near to them.

According to statistical analysis of 625 data, we get the proportion distribution in different error interval, which is presented in Fig. 7. From it we can see that the error distributions of our system are concentrate within 2.5 meters, and the error value less than 1.5 meters account for 79.37%. While the localization errors from CC2431 system less than 3 meters are tend to even distribution holding 50.4%, and the error value more than 3 meters hold 49.6% in our test environment.

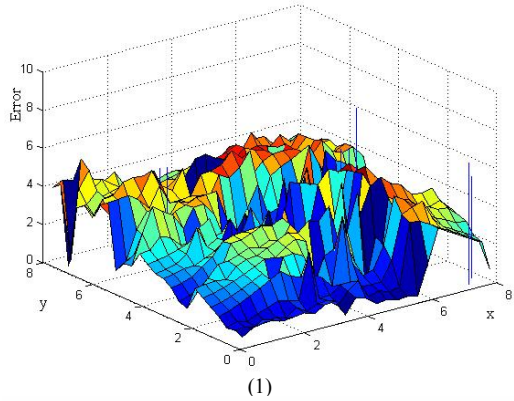

(1)

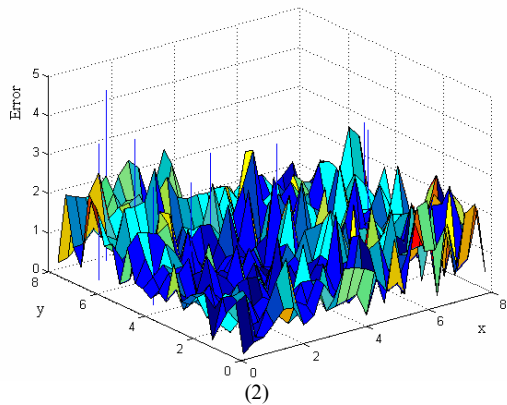

(2)

Figure 6. Error value in different area: (1) Error distribution of TI CC2431 localization system in our test environment. (2) Error distribution when using our algorithm in the same environment.

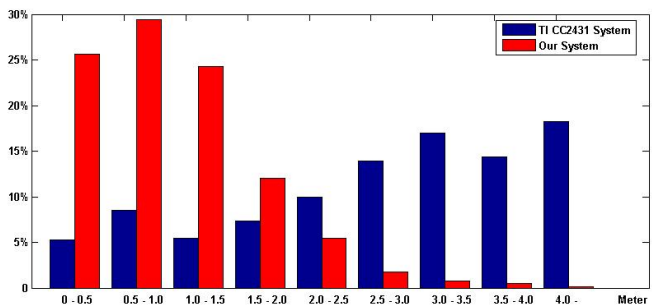

Figure 7. Error distribution of two systems.

## V. CONCLUSIONS

Localization issue is a critical and supporting technique in WSNs. Great efforts have been made by many researchers and a variant of algorithm also have been proposed. In this paper we mainly focus on the usability and availability in realistic environment and applications systems. Compared with traditional algorithm, our scheme is lightweight and succinct. We implement our algorithm and the localization results we test show that our algorithm has a better performance compared with the CC2431 system at the same test environment.

Our algorithm also has some things to be desired, for example, if the blind node moves with a high speed, the localization result descend. In addition, how to maintain nearly the same performance when lengthen the distance between two reference nodes. All these questions also existed for most other localization in WSNs. More ingenious algorithms are expected to be designed out to solve these problems. Absolutely, all these efforts will greatly accelerate the development of WSNs technique to be applied into large-scale application systems.

ACKNOWLEDGMENT

This work was partially supported by the National Natural Science Foundation of China under Grant No. 60903153, and the Fundamental Research Funds for Central Universities: DUT10ZD110.